\documentclass[twocolumn]{revtex4}
\usepackage{graphicx,epsf}
\usepackage{color}
\usepackage{dcolumn}
\usepackage{amsfonts}
\usepackage{mathrsfs}

\begin{document}


\title{High frequency mode generation by toroidal Alfv\'en eigenmodes}

\author{Shizhao Wei$^1$, Peiwan Shi$^2$,   Liming Yu$^2$, Wei Chen$^2$,  Ningfei Chen$^1$  and Zhiyong Qiu$^{1}$\footnote{Corresponding author: zqiu@zju.edu.cn}}

\affiliation{$^1$Institute for    Fusion Theory and Simulation and Department of Physics, Zhejiang University, Hangzhou, P.R.C\\
$^2$Southwestern Institute of Physics - P.O. Box 432 Chengdu 610041, P.R.C.}

\begin{abstract}
Nonlinear generation of  high frequency mode (HFM) by toroidal Alfv\'en eigenmode (TAE) observed in HL-2A tokamak   is analyzed using nonlinear gyrokinetic theory. It is found that, the HFM can be dominated by $|nq-m|=1$ perturbations with predominantly ideal magnetohydrodynamic if the two primary TAEs are co-propagating; while the HFM can be characterized by $nq-m=0$ electrostatic perturbations if the two primary TAEs are counter-propagating. Here, $n$ and $m$ are respectively the toroidal and poloidal mode numbers, and $q$ is the safety factor. The nonlinear process is sensitive to the equilibrium magnetic geometry of the device.
\end{abstract}

\maketitle

In burning plasmas of next generation magnetically confined fusion devices,     the alpha particles  related physics are  of  crucial importance for the burning plasma performance \cite{AFasoliNF2007,PLauberPR2013,LChenRMP2016}.
Alpha particles, and more generally, energetic particles (EPs) including super-thermal particles due to auxiliary heating, contribute to the heating of thermal plasmas via Coulomb collisions,
resonant  excitation of symmetry breaking shear Alfv\'en wave (SAW) instabilities \cite{MRosenbluthPRL1975,LChenVarenna1988} and the related EP loss, leading to degradation of confinement and potential  damage of plasma facing components due to the heavy heat load \cite{RDingNF2015}.
The in-depth understanding of the SAW instability, including resonant excitation by alpha particles \cite{LChenVarenna1988,GFuPoFB1989,FZoncaPoP2014b}, nonlinear dynamic evolution and saturation \cite{HBerkPoFB1990c,TSHahmPRL1995,FZoncaPRL1995,LChenPoP2013,HZhangPRL2012,JZhuNF2014,YTodoNF2010,YTodoNF2012a,YTodoNF2012b,FZoncaNJP2015,LChenRMP2016}, and the induced alpha particle transport \cite{LChenJGR1999,MSchnellerPPCF2016,ZQiuNF2019,MFalessiPoP2019}, are key elements of fusion plasma physics, and are under intensive investigation.

In tokamaks with complex  equilibrium magnetic geometry and plasma nonuniformities, SAW instabilities can be excited as various discrete Alfv\'en eigenmodes inside the forbidden gaps of SAW continuous spectrum induced by various symmetry breaking mechanisms \cite{CZChengAP1985,WHeidbrinkPRL1993},   to minimize the continuum damping and thus, excitation threshold in EP driving  intensity.
One well studied Alfv\'en eigenmode  is the famous  toroidal Alfv\'en eigenmode (TAE) \cite{CZChengAP1985}  excited inside the   SAW continuum gap  induced by the coupling of neighbouring poloidal harmonics of SAW continuum due to toroidicity. TAEs are characterized with two dominant poloidal harmonics ($m$ and $m+1$)   localized near the center of two mode rational surfaces, where $q(r)\simeq (2m+1)/(2n)$, and their characteristic parallel wavenumber both being $|k_{\parallel}|\simeq 1/(2qR_0)$. Here, $n/m$ are the toroidal/poloidal mode numbers, and $q$ is the safety factor.
TAE is recognized  as one of the most dangerous candidates in transporting EPs due to its low excitation threshold   and its optimized resonance condition with fusion alpha particles,  e.g.,   $v_{\alpha}\gtrsim V_A$  in ITER \cite{AFasoliNF2007}. Here, $v_{\alpha}$ is the birth  velocity of the $3.52MeV$ fusion alpha particle, and $V_A$ is the Alfv\'en velocity.

The nonlinear dynamics of SAW instabilities, determining their nonlinear saturation level and spectrum, are thus, of significant importance for the qualitative and quantitative  understanding of alpha particle confinement and consequently, fusion plasma performance. In  next generation tokamaks with $a/\rho_E\gg O(10)$ \cite{AFasoliNF2007}, most unstable TAEs are characterized by toroidal mode numbers $n\gtrsim O(10)$ \cite{LChenRMP2016,TWangPoP2018}. Here, $a$ is the minor radius of the torus, and $\rho_E$ is the EP characteristic orbit width. As a result,  many ($\sim O(n^2q)$) TAEs co-exist simultaneously.
Thus, the nonlinear mode  coupling processes, as a channel for TAE nonlinear dynamics, are expected to play a crucial role in burning plasmas, leading to complex nonlinear interactions and spectral transfer.
Nonlinear mode couplings of TAEs are investigated via numerical simulations \cite{DSpongPoP1994,YTodoNF2010} as well as  analytical theory \cite{FZoncaPRL1995,LChenPPCF1998,TSHahmPRL1995,LChenPRL2012,ZQiuNF2017,ZQiuPRL2018,ZQiuNF2019} for the development of  predictive ability of  burning plasma scenarios expecting to have rich   TAE nonlinear mode coupling phenomena. On the other hand, TAE related nonlinear mode couplings are rarely observed in present day tokamaks \cite{MVanZeelandPoP2007,WChenEPL2014}, probably due to the low $a/\rho_E$ value as well as the limitation of diagnosis with high frequency/radial resolution in central core plasmas.

In a recent HL-2A experiment, high frequency modes (HFM) in the ellipticity induced Alfv\'en eigenmode (EAE) frequency range was observed \cite{PShiNF2019}, characterized by toroidally symmetric ($n=0$) mode structures. In the experiment, it was observed that, due to the existence of the large amplitude magnetic island with toroidal  mode number    $n=1$, a series of TAEs with neighboring toroidal mode numbers were generated, as reported in Ref. \citenum{WChenEPL2014}. It was further found that \cite{PShiNF2019}, two TAEs may couple and generate HFM, with the frequency being double of that of TAE, i.e., in the EAE frequency range. A special case analyzed in Rf. \cite{PShiNF2019} is that,
two TAEs with opposite toroidal mode numbers, may couple and generate toroidally symmetric HFM.  Note that, though in the EAE frequency range, these nonlinearly generated HFMs are not necessarily EAE gap modes. In fact, they can exist in tokamaks with circular cross-sections, as we show in the following analysis.  The generation of the toroidally symmetric  HFM, may provide an additional channel for the primary  TAE saturation. The HL-2A observation of HFM generation due to nonlinear coupling of TAEs \cite{PShiNF2019}, also provides  a scenario for the verification of the sophisticated nonlinear theoretical approaches. In the rest of this paper,  this nonlinear coupling  process is analyzed using nonlinear gyrokinetic theory \cite{EFriemanPoF1982}.

For two TAEs to generate the   HFM with toroidal mode number denoted by $n_h$, assuming $\Omega_h(\omega_h,\mathbf{k}_h)=\Omega_1(\omega_1,\mathbf{k}_1)+\Omega_2(\omega_2,\mathbf{k}_2)$ as the wavenumber/frequency matching conditions,  the two TAEs ($\Omega_1$ and $\Omega_2$) have  $\omega_1\simeq\omega_2\simeq V_A/(2qR_0)$ and $n_1+n_2=n_h$. Note that, for $\Omega_1$ and $\Omega_2$ both with $|k_{\parallel}|\equiv |(nq-m)/(qR_0)|\simeq 1/(2qR_0)$, $|k_{\parallel,h}|=|k_{1,\parallel}+k_{2,\parallel}|= 0\ \mbox{or}\  1/(qR_0)$, corresponding to $|n_h q-m_h|$ being 0 or 1, depending on the  sign of $k_{1,\parallel}\times k_{2,\parallel}$.  In the following analysis,   these two cases are denoted  as, respectively, co- and counter- propagating cases, since the parallel phase velocities of two TAE with same frequency, defined as $\omega/k_{\parallel}$, are in the same direction (co-propagating) for $k_{1,\parallel}k_{2,\parallel}>0$ while in the opposite direction (counter-propagating) for $k_{1,\parallel}k_{2,\parallel}<0$. As a result, $|k_{h,\parallel}|=1/(qR_0)$ in the co-propagating limit, and consequently, $|n_hq-m_h|=1$. On the other hand, in the counter-propagating limit, $k_{h,\parallel}=0$ and consequently, $n_hq-m_h=0$. These two cases will be discussed separately, due to the very different particle responses. For the special case of toroidally symmetry HFM generation \cite{PShiNF2019}, one has $n_h/m_h=0/1$ or $0/0$ for the co- and counter-propagating cases, respectively.

To study the nonlinear coupling process, $\delta\phi$ and $\delta\psi\equiv \omega\delta A_{\parallel}/(ck_{\parallel})$ are adopted as the field variables, with $\delta\phi$ and $\delta A_{\parallel}$ being respectively the scalar potential and parallel component of the vector potential, and ideal MHD constraint can be forced by taking $\delta\phi=\delta\psi$.  For the simplicity of discussion and focus on the main physics picture, we assume the scalar potentials have one dominant toroidal and poloidal mode number, i.e.,
\begin{eqnarray}
\delta\phi_k=A_ke^{i(n\phi-m\theta+k_r r -\omega t)},\nonumber
\end{eqnarray}
with  $A_k$ being the mode amplitude. In the more general case, for a mode with given toroidal mode number, $\sim O(nq)$ poloidal harmonics are coupled together due to toroidicity, and one generally needs to adopt the ballooning formalism.  The governing nonlinear equations are then derived from quasi-neutrality condition:
\begin{eqnarray}
\frac{n_0e^2}{T_i}\left(1+\frac{T_i}{T_e}\right)\delta\phi_k=\sum_s \left\langle q J_k\delta H_k \right\rangle_s,\label{eq:QN}
\end{eqnarray}
and nonlinear vorticity equation \cite{LChenNF2001}
\begin{eqnarray}
&&\frac{c^2}{4\pi \omega^2_k}B\frac{\partial}{\partial l}\frac{k^2_{\perp}}{B}\frac{\partial}{\partial l}\delta \psi_k +\frac{e^2}{T_i}\left\langle (1-J^2_ k)F_0\right\rangle\delta\phi_k\nonumber\\
&=&-i\frac{c}{B_0\omega_k}\sum_{\mathbf{k}=\mathbf{k}'+\mathbf{k}''} \mathbf{\hat{b}}\cdot\mathbf{k}''\times\mathbf{k}'\left [ \frac{c^2}{4\pi}k''^2_{\perp} \frac{\partial_l\delta\psi_{k'}\partial_l\delta\psi_{k''}}{\omega_{k'}\omega_{k''}} \right.\nonumber\\
&&\left.+ \left\langle e(J_kJ_{k'}-J_{k''})\delta L_{k'}\delta H_{k''}\right\rangle \right].
\label{eq:vorticityequation}
\end{eqnarray}
Here, $J_k\equiv J_0(k_{\perp}\rho)$ with $J_0$ being the Bessel function of zero index accounting for finite Larmor radius effects, $\rho=v_{\perp}/\Omega_c$ is the particle Larmor radius, $\Omega_c=eB/(mc)$ is the cyclotron frequency, $F_0$ is the equilibrium particle distribution function,  $\sum_s$ is the summation on different particle species with $s=e,i$ denoting electrons and ions,    $\omega_d=(v^2_{\perp}+2
v^2_{\parallel})/(2 \Omega_c R_0)\left(k_r\sin\theta+k_{\theta}\cos\theta\right)$ is the magnetic drift frequency,  $l$ is the arc length along the equilibrium magnetic field line, $\delta L_k\equiv\delta\phi_k-k_{\parallel} v_{\parallel}\delta\psi_k/\omega_k$; and other notations are standard. The usual curvature coupling term accounting for   thermal plasma compressibility, and thus, beta-induced Alfv\'en eigenmode (BAE) related physics \cite{WHeidbrinkPRL1993,FZoncaPPCF1996}, are neglected since both TAEs and HFM of interest here  have frequencies much higher than BAE.
The  two terms on the right hand side of equation (\ref{eq:vorticityequation}) are the   dominant  nonlinear terms, i.e.,    Maxwell and Reynolds stresses \cite{LChenPoP2013}. We note that, Reynolds and Maxwell stresses are the dominant nonlinear term in the short wavelength kinetic regime with $k^2_{\perp}\rho^2_i\gg\omega/\Omega_{ci}$, which may not be the case in HL-2A experiments where TAEs are typically characterized by $n\sim O(1)$. However, we use the kinetic expression due to two reasons: 1. the qualitative picture for nonlinear mode coupling is not changed and 2. the obtained results can be applied to next generation tokamaks, which indeed operate  in the parameter regime where kinetic expression is relevant \cite{ZQiuNF2019}.   Furthermore, $\langle\cdots\rangle$ is the velocity space integration and $\delta H_k$ is the nonadiabatic particle response, which is derived from nonlinear gyrokinetic equation \cite{EFriemanPoF1982}:
\begin{eqnarray}
&&\left(-i\omega+v_{\parallel}\partial_l+i\omega_d\right)\delta H_k=-i\omega_k\frac{q}{T}F_0J_k\delta L_k \nonumber\\
&&\hspace*{4em}-\frac{c}{B_0}\sum_{\mathbf{k}=\mathbf{k}'+\mathbf{k}''}\mathbf{b}\cdot\mathbf{k''}\times\mathbf{k'}J_{k'}\delta L_{k'}\delta H_{k''}\label{eq:NLGKE}.
\end{eqnarray}
In equation (\ref{eq:NLGKE}), the   free energy associated with pressure gradient is neglected, assuming thermal plasmas dominating the nonlinear mode coupling process don't contribute to drive the primary TAEs unstable.  A small amplitude expansion is taken, i.e., separating $\delta H_k=\delta H^L_k+\delta H^{NL}_k$, with the subscripts ``L" and ``NL" denoting linear and nonlinear particle responses, and the particle responses are derived from equation (\ref{eq:NLGKE}) order by order.

For TAEs with $|k_{\parallel}v_{t,e}|\gg|\omega_T|\gg |k_{\parallel}v_{t,i}|,|\omega_d|$, the linear particle responses can be readily derived as $\delta H^L_{T,i}=(e/T_i)F_0J_k\delta\phi_k$ and $\delta H^L_{T,e}=-(e/T_e)F_0\delta\psi_k$, which are used to derive the nonlinear particle responses to the HFM. Since the particle responses to the HFMs generated by co-/counter- propagating TAEs are very different, the two cases will be investigated separately.

{\bf{Co-propagating TAEs.}} For $\Omega_1$ and $\Omega_2$ being co-propagating,  one then has, $|k_{h,\parallel}|=1/(qR_0)$, $k_{h,\parallel}v_{t,e}\gg\omega_h\simeq V_A/(qR_)\gg k_{h,\parallel}v_{t,i},\omega_d$, and consequently, $\delta H^L_{h,i}=(e/T_e)F_0J_h\delta\phi_h$ and $\delta H^L_{h,e}=-(e/T_e)F_0\delta\psi_h$.  The nonlinear electron response to $\Omega_h$, can thus, be derived from
\begin{eqnarray}
v_{\parallel}\partial_l \delta H^{NL}_{h,e}&=&-\frac{c}{B_0}\sum\mathbf{\hat{b}}\cdot\mathbf{k''}\times\mathbf{k'} \delta L_{k'}\delta H_{k'',e}\nonumber\\
&\simeq&-\frac{c}{B_0}v_{\parallel}\frac{e}{T_e}\mathbf{\hat{b}}\cdot\mathbf{k_1}\times\mathbf{k_2} \left(\frac{k_{2,\parallel}}{\omega_2}-\frac{k_{1,\parallel}}{\omega_1}\right)\delta\psi_1\delta\psi_2.\nonumber
\end{eqnarray}
Finite $\delta H^{NL}_{h,e}$ comes from the $\omega_T$ dependence on the toroidal mode number, and can be neglected. On the other hand, the nonlinear ion response to $\Omega_h$, can be derived from
\begin{eqnarray}
-i\omega_h\delta H^{NL}_{h,i}=-\frac{c}{B_0}\sum\mathbf{\hat{b}}\cdot\mathbf{k''}\times\mathbf{k'} J_{k'}\delta L_{k'}\delta H_{k'',i},\nonumber
\end{eqnarray}
which also has negligible contribution due to the weak dependence of $\omega_T$ on $k_{\perp}\rho_i$. As a result, one then has, from the quasi-neutrality condition, $\delta\phi_h=\delta\psi_h$ to the leading order, i.e., the ideal MHD constraint of $\Omega_h$ is not broken by nonlinear effects.

The nonlinear $\Omega_h$ dispersion relation, can then be derived from the nonlinear vorticity equation, and one has
\begin{eqnarray}
&&\frac{c^2B_0}{4\pi\omega^2_h} \frac{\partial}{\partial l}\frac{k^2_{h,\perp}}{B_0}\frac{\partial}{\partial l}\delta\psi_h +\frac{e^2}{Ti}\left\langle (1-J^2_h)F_0\right\rangle\delta\phi_h\nonumber\\
&\simeq& i\frac{c}{B\omega_h}\hat{\mathbf{b}}\cdot\mathbf{k_2}\times\mathbf{k_1} \left[\frac{c^2}{4\pi}\left(k^2_{2,\perp}-k^2_{1,\perp}\right)\frac{k_{1,\parallel}k_{2,\parallel}\delta\phi_1\delta\phi_2}{\omega_1\omega_2}\right.\nonumber\\
&&\left.-\langle e(J_1-J_2)\left(\delta\phi_2\delta H_{2,i}+\delta\phi_1\delta H_{1,i}\right)\rangle\right].\label{eq:NL_co_vorticity}
\end{eqnarray}
Substituting the linear particle responses into equation (\ref{eq:NL_co_vorticity}),  noting that $k_{1,\theta}=-k_{2,\theta}$,  $J^2(k_{\perp}\rho_i)\simeq 1-k^2_{\perp}\rho^2_i/4$, $k_{h,r}=k_{1,r}+k_{2,r}$ and that for TAEs, $k_r\sim k_{\theta}/\epsilon_0$, one obtains
\begin{eqnarray}
&&(1-k^2_{h,\parallel}V^2_A/\omega^2_h)\delta\phi_h\nonumber\\
&=&  \frac{c}{B_0\omega_h} \left(1-\frac{k_{1,\parallel}k_{2,\parallel}V^2_A}{\omega_1\omega_2}\right) \left(\delta\phi_2\partial_r\delta\phi_1-\delta\phi_1\partial_r\delta\phi_2\right).\nonumber
\end{eqnarray}
Noting that $|k_{h,\parallel}|\simeq 2|k_{1,\parallel}|\simeq 2|k_{2,\parallel}|$, and that $\omega_h\simeq 2\omega_1\simeq2\omega_2$, one has, $1-k^2_{h,\parallel}V^2_A/\omega^2_h\simeq 1- k_{1,\parallel}k_{2,\parallel}V^2_A/(\omega_1\omega_2)$. On the other hand, finite generation of HFM is the result of $1- k_{1,\parallel}k_{2,\parallel}V^2_A/(\omega_1\omega_2)\sim O(\epsilon)$, i.e., the breaking of the pure Alfv\'enic state by toroidicity, as discussed in Refs. \cite{LChenPRL2012,LChenPoP2013,ZQiuEPL2013}.
One then obtains, the expression of the nonlinear generated HFM:
\begin{eqnarray}
\delta\psi_h=\delta\phi_h=-\frac{c}{B_0}\frac{k_{1,\theta}}{2\omega_1}\left(\delta\phi_2\partial_r\delta\phi_1-\delta\phi_1\partial_r\delta\phi_2\right).
\end{eqnarray}

The condition $1-k^2_{h,\parallel}V^2_A/\omega^2_h\simeq 1- k_{1,\parallel}k_{2,\parallel}V^2_A/(\omega_1\omega_2)$ indicates that,  the deviation of $\omega_h$ from the center of the EAE gap   frequency ($V_A/(qR_0)$, if EAE gap exists due to finite ellipticity of the torus), is the result of $\omega_1$ and $\omega_2$ deviation from the center of the TAE gap frequency, while independent of the ellipticity. Thus, the nonlinearly generated HFM can be an EAE gap mode, depending on the relative width of ellipticity induced SAW continuum gap and the (typically downward) deviation  of $\omega_1$ and $\omega_2$ from the center of toroidicity induced SAW continuum frequency gap, due to, e.g., nonperturbative EP drive.  If the ellipticity induced SAW continuum gap is wide enough, $\Omega_h$ is an EAE gap mode which can exist without being heavily damped; while if the other limit applies, $\Omega_h$ will strongly couple to SAW continuum and be heavily continuum damped, leading to the nonlinear saturation of the primary TAEs. Thus, this nonlinear process depends on the relative effects of ellipticity and toroidicity, and is sensitive to the equilibrium magnetic geometry of the torus. For HL-2A with predominantly circular geometry, we expect the HFM may couple to the SAW continuum, and is heavily damped.  Note that, in Refs. \citenum{FZoncaPRL1995,LChenPPCF1998}, the nonlinear saturation of TAE due to the nonlinear modification of the SAW continuum structure is analyzed, where two counter-propagating TAEs (with same $k_{\parallel}$ and opposite frequency) couple and generate $n/m=0/1$ low frequency magnetic field/density perturbation was analyzed.

{\bf{Counter-propagating TAEs.}} In the $\Omega_1/\Omega_2$ counter-propagating case, $\Omega_{\hbar}$ is characterized by $n_hq-m_h=0$, $\omega_{\hbar}\simeq V_A/(qR_0)$ and $k_{\hbar,\parallel}=0$. Here, subscript  $\hbar$ is used for the HFM generated by counter-propagating TAEs, to differentiate with the co-propagating case discussed above. The linear particle responses to $\Omega_{\hbar}$ can thus, be derived noting the $|\omega_{\hbar}|\gg|\omega_{tr,i}|,|\omega_{d,i}|$, and $|\omega_{tr,e}|\gg|\omega_{\hbar}|\gg|\omega_{d,e}|$ ordering. One then obtains,
$\delta H^L_{\hbar,i}=(e/T_i)F_0J_{\hbar}\delta\phi_{\hbar}$ and $\delta H^L_{\hbar,e}=-(e/T_e)F_0\overline{\delta\phi_{\hbar}}$, with $\overline{(\cdots)}\equiv (1/2\pi)\int^{2\pi}_0d\theta(\cdots)$ denoting surface averaging.  The nonlinear vorticity equation of $\Omega_{\hbar}$ can be derived as
\begin{eqnarray}
&&\frac{e^2}{T_i}\langle(1-J^2_{\hbar})F_0\rangle\delta\phi_{\hbar}\nonumber\\
&\simeq& -i\frac{c}{B_0\omega_{\hbar}}\hat{\mathbf{b}}\cdot\mathbf{k_1}\times\mathbf{k_2}\left[\frac{c^2}{4\pi}(k^2_{1,\perp}-k^2_{2,\perp})\frac{-k_{1,\parallel}k_{2,\parallel}}{\omega_1\omega_2}\right.\nonumber\\
&&\left.+\langle e(J_2-J_1)(\delta\phi_2\delta H_{1,i}+\delta\phi_1\delta H_{2,i})\rangle\right]\nonumber\\
&\simeq&    - \frac{c}{2B_0\omega_h}\frac{n_0e^2}{T_i} \rho^2_{t,i}k_{1,\theta} \left[1- k_{1,\parallel}k_{2,\parallel}V^2_A/(\omega_1\omega_2)\right]\nonumber\\ &&\times\partial^2_r(\delta\phi_2\partial_r\delta\phi_1-\delta\phi_1\partial_r\delta\phi_2).\label{eq:NL_counter_vorticity}
\end{eqnarray}
In deriving equation (\ref{eq:NL_counter_vorticity}),   the field line bending and curvature coupling terms   of the nonlinear vorticity equation are neglected, in consistency with $k_{\hbar,\parallel}=0$ and that $\Omega_{\hbar}$ frequency is much higher than BAE frequency. Noting again that $k_{h,r}=k_{1,r}+k_{2,r}$, $k_r\simeq k_{\theta}/\epsilon$ for TAEs in the inertial layer where nonlinear coupling dominates, and that $k_{1,\parallel}/\omega_1\simeq -k_{2,\parallel}/\omega_2$, one obtains
\begin{eqnarray}
\delta\phi_{\hbar}=\frac{c}{B_0}\frac{k_{1,\theta}}{\omega_1}\left(\delta\phi_2\partial_r\delta\phi_1-\delta\phi_1\partial_r\delta\phi_2\right).
\end{eqnarray}
In this case, the HFM is an electrostatic perturbation, with toroidally and poloidally symmetric (zonal) mode structures. The  electromagnetic component of $\Omega_{\hbar}$, can only enter through curvature coupling term, and is dominated by $m_h+2$ poloidal harmonic, similar to the electromagnetic component of geodesic acoustic mode \cite{ASmolyakovPPCF2008}. Inclusion of the electromagnetic component, will induce  $O(k^2_{\perp}\rho^2_i)$ corrections to the present result, and is neglected. The   nonlinear  Reynolds and Maxwell stresses due to the coupling of the two counter-propagating TAEs, are additive, as discussed in Ref. \cite{ZQiuJPSCP2014} where $n/m=0/1$ ion sound wave with the frequency much lower than TAE frequency is generated by two counter-propagating TAEs (with same $k_{\parallel}$ while opposite frequency as in Ref. \cite{LChenPPCF1998,FZoncaPRL1995}) as one mechanism for TAE nonlinear saturation. The impact of the  $n_hq-m_h=0$ HFM on prime TAE nonlinear dynamics including saturation, will be investigated both experimentally and theoretically  in a future publication.

In conclusion, the   toroidally symmetric high frequency mode (HFM) generation by TAEs observed in HL-2A experiment is analyzed using nonlinear gyrokinetic theory. It is found that, two different cases may exist, i.e., the TAEs be co- or counter- propagating. In the co-propagating case, the two TAEs have the same parallel wavenumber, and the generated HFM, is characterized by  $|n_hq-m_h|=1$ harmonic, and    ideal MHD character ($\delta E_{\parallel}=0$) to the leading order. The generated HFM mode, can be a weakly damped EAE in the ellipticity induced SAW continuum gap, or a heavily damped mode, depending on the relative role of ellipticity and toroidicity. Thus, this discussed process strongly depends on the equilibrium magnetic geometry of the tokamak.  In the counter-propagating case, the two TAEs have opposite parallel wavenumber, and the generated HFM  is characterized by $|n_hq-m_h|=0$. The HFM is thus, predominantly electrostatic. The corresponding HFM amplitude in both cases are derived. The detailed comparison with experimental observations, and the consequence on TAE nonlinear dynamics, will be reported in a future publication.

This work is supported by   the National Key R\&D Program of China  under Grant Nos.  2017YFE0301900 and 2017YFE0301202,
the National Science Foundation of China under grant Nos.  11575157, 11875024 and 11875233,  and Fundamental Research
Fund for Chinese Central Universities.


\begin{thebibliography}{39}
\expandafter\ifx\csname natexlab\endcsname\relax\def\natexlab#1{#1}\fi
\expandafter\ifx\csname bibnamefont\endcsname\relax
  \def\bibnamefont#1{#1}\fi
\expandafter\ifx\csname bibfnamefont\endcsname\relax
  \def\bibfnamefont#1{#1}\fi
\expandafter\ifx\csname citenamefont\endcsname\relax
  \def\citenamefont#1{#1}\fi
\expandafter\ifx\csname url\endcsname\relax
  \def\url#1{\texttt{#1}}\fi
\expandafter\ifx\csname urlprefix\endcsname\relax\def\urlprefix{URL }\fi
\providecommand{\bibinfo}[2]{#2}
\providecommand{\eprint}[2][]{\url{#2}}

\bibitem[{\citenamefont{Fasoli et~al.}(2007)\citenamefont{Fasoli, Gormenzano,
  Berk, Breizman, Briguglio, Darrow, Gorelenkov, Heidbrink, Jaun, Konovalov
  et~al.}}]{AFasoliNF2007}
\bibinfo{author}{\bibfnamefont{A.}~\bibnamefont{Fasoli}},
  \bibinfo{author}{\bibfnamefont{C.}~\bibnamefont{Gormenzano}},
  \bibinfo{author}{\bibfnamefont{H.}~\bibnamefont{Berk}},
  \bibinfo{author}{\bibfnamefont{B.}~\bibnamefont{Breizman}},
  \bibinfo{author}{\bibfnamefont{S.}~\bibnamefont{Briguglio}},
  \bibinfo{author}{\bibfnamefont{D.}~\bibnamefont{Darrow}},
  \bibinfo{author}{\bibfnamefont{N.}~\bibnamefont{Gorelenkov}},
  \bibinfo{author}{\bibfnamefont{W.}~\bibnamefont{Heidbrink}},
  \bibinfo{author}{\bibfnamefont{A.}~\bibnamefont{Jaun}},
  \bibinfo{author}{\bibfnamefont{S.}~\bibnamefont{Konovalov}},
  \bibnamefont{et~al.}, \bibinfo{journal}{Nuclear Fusion}
  \textbf{\bibinfo{volume}{47}}, \bibinfo{pages}{S264} (\bibinfo{year}{2007}).

\bibitem[{\citenamefont{Lauber}(2013)}]{PLauberPR2013}
\bibinfo{author}{\bibfnamefont{P.}~\bibnamefont{Lauber}},
  \bibinfo{journal}{Physics Reports} \textbf{\bibinfo{volume}{533}},
  \bibinfo{pages}{33 } (\bibinfo{year}{2013}), ISSN \bibinfo{issn}{0370-1573}.

\bibitem[{\citenamefont{Chen and Zonca}(2016)}]{LChenRMP2016}
\bibinfo{author}{\bibfnamefont{L.}~\bibnamefont{Chen}} \bibnamefont{and}
  \bibinfo{author}{\bibfnamefont{F.}~\bibnamefont{Zonca}},
  \bibinfo{journal}{Review of Modern Physics} \textbf{\bibinfo{volume}{88}},
  \bibinfo{pages}{015008} (\bibinfo{year}{2016}).

\bibitem[{\citenamefont{Rosenbluth and Rutherford}(1975)}]{MRosenbluthPRL1975}
\bibinfo{author}{\bibfnamefont{M.}~\bibnamefont{Rosenbluth}} \bibnamefont{and}
  \bibinfo{author}{\bibfnamefont{P.}~\bibnamefont{Rutherford}},
  \bibinfo{journal}{Phys. Rev. Lett.} \textbf{\bibinfo{volume}{34}},
  \bibinfo{pages}{1428} (\bibinfo{year}{1975}).

\bibitem[{\citenamefont{Chen}(1988)}]{LChenVarenna1988}
\bibinfo{author}{\bibfnamefont{L.}~\bibnamefont{Chen}}, in
  \emph{\bibinfo{booktitle}{Theory of Fusion Plasmas}}, edited by
  \bibinfo{editor}{\bibfnamefont{J.}~\bibnamefont{Vaclavik}},
  \bibinfo{editor}{\bibfnamefont{F.}~\bibnamefont{Troyon}}, \bibnamefont{and}
  \bibinfo{editor}{\bibfnamefont{E.}~\bibnamefont{Sindoni}}
  (\bibinfo{publisher}{Association EUROATOM, Bologna}, \bibinfo{year}{1988}),
  p. \bibinfo{pages}{327}.

\bibitem[{\citenamefont{Ding et~al.}(2015)\citenamefont{Ding, Pitts, Borodin,
  Carpentier, Ding, Gong, Guo, Kirschner, Kocan, Li et~al.}}]{RDingNF2015}
\bibinfo{author}{\bibfnamefont{R.}~\bibnamefont{Ding}},
  \bibinfo{author}{\bibfnamefont{R.}~\bibnamefont{Pitts}},
  \bibinfo{author}{\bibfnamefont{D.}~\bibnamefont{Borodin}},
  \bibinfo{author}{\bibfnamefont{S.}~\bibnamefont{Carpentier}},
  \bibinfo{author}{\bibfnamefont{F.}~\bibnamefont{Ding}},
  \bibinfo{author}{\bibfnamefont{X.}~\bibnamefont{Gong}},
  \bibinfo{author}{\bibfnamefont{H.}~\bibnamefont{Guo}},
  \bibinfo{author}{\bibfnamefont{A.}~\bibnamefont{Kirschner}},
  \bibinfo{author}{\bibfnamefont{M.}~\bibnamefont{Kocan}},
  \bibinfo{author}{\bibfnamefont{J.}~\bibnamefont{Li}}, \bibnamefont{et~al.},
  \bibinfo{journal}{Nuclear Fusion} \textbf{\bibinfo{volume}{55}},
  \bibinfo{pages}{023013} (\bibinfo{year}{2015}).

\bibitem[{\citenamefont{Fu and Van~Dam}(1989)}]{GFuPoFB1989}
\bibinfo{author}{\bibfnamefont{G.~Y.} \bibnamefont{Fu}} \bibnamefont{and}
  \bibinfo{author}{\bibfnamefont{J.~W.} \bibnamefont{Van~Dam}},
  \bibinfo{journal}{Physics of Fluids B} \textbf{\bibinfo{volume}{1}},
  \bibinfo{pages}{1949} (\bibinfo{year}{1989}).

\bibitem[{\citenamefont{Zonca and Chen}(2014)}]{FZoncaPoP2014b}
\bibinfo{author}{\bibfnamefont{F.}~\bibnamefont{Zonca}} \bibnamefont{and}
  \bibinfo{author}{\bibfnamefont{L.}~\bibnamefont{Chen}},
  \bibinfo{journal}{Physics of Plasmas} \textbf{\bibinfo{volume}{21}},
  \bibinfo{eid}{072121} (\bibinfo{year}{2014}).

\bibitem[{\citenamefont{Berk and Breizman}(1990)}]{HBerkPoFB1990c}
\bibinfo{author}{\bibfnamefont{H.~L.} \bibnamefont{Berk}} \bibnamefont{and}
  \bibinfo{author}{\bibfnamefont{B.~N.} \bibnamefont{Breizman}},
  \bibinfo{journal}{Physics of Fluids B} \textbf{\bibinfo{volume}{2}},
  \bibinfo{pages}{2246} (\bibinfo{year}{1990}).

\bibitem[{\citenamefont{Hahm and Chen}(1995)}]{TSHahmPRL1995}
\bibinfo{author}{\bibfnamefont{T.~S.} \bibnamefont{Hahm}} \bibnamefont{and}
  \bibinfo{author}{\bibfnamefont{L.}~\bibnamefont{Chen}},
  \bibinfo{journal}{Phys. Rev. Lett.} \textbf{\bibinfo{volume}{74}},
  \bibinfo{pages}{266} (\bibinfo{year}{1995}).

\bibitem[{\citenamefont{Zonca et~al.}(1995)\citenamefont{Zonca, Romanelli,
  Vlad, and Kar}}]{FZoncaPRL1995}
\bibinfo{author}{\bibfnamefont{F.}~\bibnamefont{Zonca}},
  \bibinfo{author}{\bibfnamefont{F.}~\bibnamefont{Romanelli}},
  \bibinfo{author}{\bibfnamefont{G.}~\bibnamefont{Vlad}}, \bibnamefont{and}
  \bibinfo{author}{\bibfnamefont{C.}~\bibnamefont{Kar}},
  \bibinfo{journal}{Phys. Rev. Lett.} \textbf{\bibinfo{volume}{74}},
  \bibinfo{pages}{698} (\bibinfo{year}{1995}).

\bibitem[{\citenamefont{Chen and Zonca}(2013)}]{LChenPoP2013}
\bibinfo{author}{\bibfnamefont{L.}~\bibnamefont{Chen}} \bibnamefont{and}
  \bibinfo{author}{\bibfnamefont{F.}~\bibnamefont{Zonca}},
  \bibinfo{journal}{Physics of Plasmas} \textbf{\bibinfo{volume}{20}},
  \bibinfo{eid}{055402} (\bibinfo{year}{2013}).

\bibitem[{\citenamefont{Zhang et~al.}(2012)\citenamefont{Zhang, Lin, and
  Holod}}]{HZhangPRL2012}
\bibinfo{author}{\bibfnamefont{H.~S.} \bibnamefont{Zhang}},
  \bibinfo{author}{\bibfnamefont{Z.}~\bibnamefont{Lin}}, \bibnamefont{and}
  \bibinfo{author}{\bibfnamefont{I.}~\bibnamefont{Holod}},
  \bibinfo{journal}{Phys. Rev. Lett.} \textbf{\bibinfo{volume}{109}},
  \bibinfo{pages}{025001} (\bibinfo{year}{2012}).

\bibitem[{\citenamefont{Zhu et~al.}(2014)\citenamefont{Zhu, Ma, and
  Fu}}]{JZhuNF2014}
\bibinfo{author}{\bibfnamefont{J.}~\bibnamefont{Zhu}},
  \bibinfo{author}{\bibfnamefont{Z.}~\bibnamefont{Ma}}, \bibnamefont{and}
  \bibinfo{author}{\bibfnamefont{G.}~\bibnamefont{Fu}},
  \bibinfo{journal}{Nuclear Fusion} \textbf{\bibinfo{volume}{54}},
  \bibinfo{pages}{123020} (\bibinfo{year}{2014}).

\bibitem[{\citenamefont{Todo et~al.}(2010)\citenamefont{Todo, Berk, and
  Breizman}}]{YTodoNF2010}
\bibinfo{author}{\bibfnamefont{Y.}~\bibnamefont{Todo}},
  \bibinfo{author}{\bibfnamefont{H.}~\bibnamefont{Berk}}, \bibnamefont{and}
  \bibinfo{author}{\bibfnamefont{B.}~\bibnamefont{Breizman}},
  \bibinfo{journal}{Nuclear Fusion} \textbf{\bibinfo{volume}{50}},
  \bibinfo{pages}{084016} (\bibinfo{year}{2010}).

\bibitem[{\citenamefont{Todo et~al.}(2012)\citenamefont{Todo, Berk, and
  Breizman}}]{YTodoNF2012a}
\bibinfo{author}{\bibfnamefont{Y.}~\bibnamefont{Todo}},
  \bibinfo{author}{\bibfnamefont{H.}~\bibnamefont{Berk}}, \bibnamefont{and}
  \bibinfo{author}{\bibfnamefont{B.}~\bibnamefont{Breizman}},
  \bibinfo{journal}{Nuclear Fusion} \textbf{\bibinfo{volume}{52}},
  \bibinfo{pages}{033003} (\bibinfo{year}{2012}).

\bibitem[{\citenamefont{Todo et~al.}({2012})\citenamefont{Todo, Berk, and
  Breizman}}]{YTodoNF2012b}
\bibinfo{author}{\bibfnamefont{Y.}~\bibnamefont{Todo}},
  \bibinfo{author}{\bibfnamefont{H.~L.} \bibnamefont{Berk}}, \bibnamefont{and}
  \bibinfo{author}{\bibfnamefont{B.~N.} \bibnamefont{Breizman}},
  \bibinfo{journal}{{Nuclear Fusion}} \textbf{\bibinfo{volume}{{52}}},
  \bibinfo{pages}{094018} (\bibinfo{year}{{2012}}).

\bibitem[{\citenamefont{Zonca et~al.}(2015)\citenamefont{Zonca, Chen,
  Briguglio, Fogaccia, Vlad, and Wang}}]{FZoncaNJP2015}
\bibinfo{author}{\bibfnamefont{F.}~\bibnamefont{Zonca}},
  \bibinfo{author}{\bibfnamefont{L.}~\bibnamefont{Chen}},
  \bibinfo{author}{\bibfnamefont{S.}~\bibnamefont{Briguglio}},
  \bibinfo{author}{\bibfnamefont{G.}~\bibnamefont{Fogaccia}},
  \bibinfo{author}{\bibfnamefont{G.}~\bibnamefont{Vlad}}, \bibnamefont{and}
  \bibinfo{author}{\bibfnamefont{X.}~\bibnamefont{Wang}}, \bibinfo{journal}{New
  Journal of Physics} \textbf{\bibinfo{volume}{17}}, \bibinfo{pages}{013052}
  (\bibinfo{year}{2015}).

\bibitem[{\citenamefont{Chen}(1999)}]{LChenJGR1999}
\bibinfo{author}{\bibfnamefont{L.}~\bibnamefont{Chen}},
  \bibinfo{journal}{Journal of Geophysical Research: Space Physics}
  \textbf{\bibinfo{volume}{104}}, \bibinfo{pages}{2421} (\bibinfo{year}{1999}),
  ISSN \bibinfo{issn}{2156-2202}.

\bibitem[{\citenamefont{Schneller et~al.}(2016)\citenamefont{Schneller, Lauber,
  and Briguglio}}]{MSchnellerPPCF2016}
\bibinfo{author}{\bibfnamefont{M.}~\bibnamefont{Schneller}},
  \bibinfo{author}{\bibfnamefont{P.}~\bibnamefont{Lauber}}, \bibnamefont{and}
  \bibinfo{author}{\bibfnamefont{S.}~\bibnamefont{Briguglio}},
  \bibinfo{journal}{Plasma Physics and Controlled Fusion}
  \textbf{\bibinfo{volume}{58}}, \bibinfo{pages}{014019}
  (\bibinfo{year}{2016}).

\bibitem[{\citenamefont{Qiu et~al.}(2019)\citenamefont{Qiu, Chen, and
  Zonca}}]{ZQiuNF2019}
\bibinfo{author}{\bibfnamefont{Z.}~\bibnamefont{Qiu}},
  \bibinfo{author}{\bibfnamefont{L.}~\bibnamefont{Chen}}, \bibnamefont{and}
  \bibinfo{author}{\bibfnamefont{F.}~\bibnamefont{Zonca}},
  \emph{\bibinfo{title}{Gyrokinetic theory of the nonlinear saturation of
  toroidal alfv\'en eigenmode}}, \bibinfo{journal}{accepted by Nuclear Fusion}  (\bibinfo{year}{2019}).

\bibitem[{\citenamefont{Falessi and Zonca}(2019)}]{MFalessiPoP2019}
\bibinfo{author}{\bibfnamefont{M.~V.} \bibnamefont{Falessi}} \bibnamefont{and}
  \bibinfo{author}{\bibfnamefont{F.}~\bibnamefont{Zonca}},
  \bibinfo{journal}{Physics of Plasmas} \textbf{\bibinfo{volume}{26}},
  \bibinfo{pages}{022305} (\bibinfo{year}{2019}).

\bibitem[{\citenamefont{Cheng et~al.}(1985)\citenamefont{Cheng, Chen, and
  Chance}}]{CZChengAP1985}
\bibinfo{author}{\bibfnamefont{C.}~\bibnamefont{Cheng}},
  \bibinfo{author}{\bibfnamefont{L.}~\bibnamefont{Chen}}, \bibnamefont{and}
  \bibinfo{author}{\bibfnamefont{M.}~\bibnamefont{Chance}},
  \bibinfo{journal}{Ann. Phys.} \textbf{\bibinfo{volume}{161}},
  \bibinfo{pages}{21} (\bibinfo{year}{1985}).

\bibitem[{\citenamefont{Heidbrink et~al.}(1993)\citenamefont{Heidbrink, Strait,
  Chu, and Turnbull}}]{WHeidbrinkPRL1993}
\bibinfo{author}{\bibfnamefont{W.}~\bibnamefont{Heidbrink}},
  \bibinfo{author}{\bibfnamefont{E.}~\bibnamefont{Strait}},
  \bibinfo{author}{\bibfnamefont{M.}~\bibnamefont{Chu}}, \bibnamefont{and}
  \bibinfo{author}{\bibfnamefont{A.}~\bibnamefont{Turnbull}},
  \bibinfo{journal}{Phys. Rev. Lett.} \textbf{\bibinfo{volume}{71}},
  \bibinfo{pages}{855} (\bibinfo{year}{1993}).

\bibitem[{\citenamefont{Wang et~al.}(2018)\citenamefont{Wang, Qiu, Zonca,
  Briguglio, Fogaccia, Vlad, and Wang}}]{TWangPoP2018}
\bibinfo{author}{\bibfnamefont{T.}~\bibnamefont{Wang}},
  \bibinfo{author}{\bibfnamefont{Z.}~\bibnamefont{Qiu}},
  \bibinfo{author}{\bibfnamefont{F.}~\bibnamefont{Zonca}},
  \bibinfo{author}{\bibfnamefont{S.}~\bibnamefont{Briguglio}},
  \bibinfo{author}{\bibfnamefont{G.}~\bibnamefont{Fogaccia}},
  \bibinfo{author}{\bibfnamefont{G.}~\bibnamefont{Vlad}}, \bibnamefont{and}
  \bibinfo{author}{\bibfnamefont{X.}~\bibnamefont{Wang}},
  \bibinfo{journal}{Physics of Plasmas} \textbf{\bibinfo{volume}{25}},
  \bibinfo{pages}{062509} (\bibinfo{year}{2018}).

\bibitem[{\citenamefont{Spong et~al.}(1994)\citenamefont{Spong, Carreras, and
  Hedrick}}]{DSpongPoP1994}
\bibinfo{author}{\bibfnamefont{D.}~\bibnamefont{Spong}},
  \bibinfo{author}{\bibfnamefont{B.}~\bibnamefont{Carreras}}, \bibnamefont{and}
  \bibinfo{author}{\bibfnamefont{C.}~\bibnamefont{Hedrick}},
  \bibinfo{journal}{Physics of plasmas} \textbf{\bibinfo{volume}{1}},
  \bibinfo{pages}{1503} (\bibinfo{year}{1994}).

\bibitem[{\citenamefont{Chen et~al.}(1998)\citenamefont{Chen, Zonca, Santoro,
  and Hu}}]{LChenPPCF1998}
\bibinfo{author}{\bibfnamefont{L.}~\bibnamefont{Chen}},
  \bibinfo{author}{\bibfnamefont{F.}~\bibnamefont{Zonca}},
  \bibinfo{author}{\bibfnamefont{R.}~\bibnamefont{Santoro}}, \bibnamefont{and}
  \bibinfo{author}{\bibfnamefont{G.}~\bibnamefont{Hu}},
  \bibinfo{journal}{Plasma physics and controlled fusion}
  \textbf{\bibinfo{volume}{40}}, \bibinfo{pages}{1823} (\bibinfo{year}{1998}).

\bibitem[{\citenamefont{Chen and Zonca}(2012)}]{LChenPRL2012}
\bibinfo{author}{\bibfnamefont{L.}~\bibnamefont{Chen}} \bibnamefont{and}
  \bibinfo{author}{\bibfnamefont{F.}~\bibnamefont{Zonca}},
  \bibinfo{journal}{Phys. Rev. Lett.} \textbf{\bibinfo{volume}{109}},
  \bibinfo{pages}{145002} (\bibinfo{year}{2012}).

\bibitem[{\citenamefont{Qiu et~al.}(2017)\citenamefont{Qiu, Chen, and
  Zonca}}]{ZQiuNF2017}
\bibinfo{author}{\bibfnamefont{Z.}~\bibnamefont{Qiu}},
  \bibinfo{author}{\bibfnamefont{L.}~\bibnamefont{Chen}}, \bibnamefont{and}
  \bibinfo{author}{\bibfnamefont{F.}~\bibnamefont{Zonca}},
  \bibinfo{journal}{Nuclear Fusion} \textbf{\bibinfo{volume}{57}},
  \bibinfo{pages}{056017} (\bibinfo{year}{2017}).

\bibitem[{\citenamefont{Qiu et~al.}(2018)\citenamefont{Qiu, Chen, Zonca, and
  Chen}}]{ZQiuPRL2018}
\bibinfo{author}{\bibfnamefont{Z.}~\bibnamefont{Qiu}},
  \bibinfo{author}{\bibfnamefont{L.}~\bibnamefont{Chen}},
  \bibinfo{author}{\bibfnamefont{F.}~\bibnamefont{Zonca}}, \bibnamefont{and}
  \bibinfo{author}{\bibfnamefont{W.}~\bibnamefont{Chen}},
  \bibinfo{journal}{Phys. Rev. Lett.} \textbf{\bibinfo{volume}{120}},
  \bibinfo{pages}{135001} (\bibinfo{year}{2018}).

\bibitem[{\citenamefont{Van~Zeeland et~al.}(2007)\citenamefont{Van~Zeeland,
  Austin, Gorelenkov, Heidbrink, Kramer, Makowski, McKee, Nazikian, Ruskov, and
  Turnbull}}]{MVanZeelandPoP2007}
\bibinfo{author}{\bibfnamefont{M.~A.} \bibnamefont{Van~Zeeland}},
  \bibinfo{author}{\bibfnamefont{M.~E.} \bibnamefont{Austin}},
  \bibinfo{author}{\bibfnamefont{N.~N.} \bibnamefont{Gorelenkov}},
  \bibinfo{author}{\bibfnamefont{W.~W.} \bibnamefont{Heidbrink}},
  \bibinfo{author}{\bibfnamefont{G.~J.} \bibnamefont{Kramer}},
  \bibinfo{author}{\bibfnamefont{M.~A.} \bibnamefont{Makowski}},
  \bibinfo{author}{\bibfnamefont{G.~R.} \bibnamefont{McKee}},
  \bibinfo{author}{\bibfnamefont{R.}~\bibnamefont{Nazikian}},
  \bibinfo{author}{\bibfnamefont{E.}~\bibnamefont{Ruskov}}, \bibnamefont{and}
  \bibinfo{author}{\bibfnamefont{A.~D.} \bibnamefont{Turnbull}},
  \bibinfo{journal}{Physics of Plasmas} \textbf{\bibinfo{volume}{14}},
  \bibinfo{pages}{056102} (\bibinfo{year}{2007}).

\bibitem[{\citenamefont{Chen et~al.}(2014)\citenamefont{Chen, Qiu, Ding, Xie,
  Yu, Ji, Li, Li, Dong, Shi et~al.}}]{WChenEPL2014}
\bibinfo{author}{\bibfnamefont{W.}~\bibnamefont{Chen}},
  \bibinfo{author}{\bibfnamefont{Z.}~\bibnamefont{Qiu}},
  \bibinfo{author}{\bibfnamefont{X.}~\bibnamefont{Ding}},
  \bibinfo{author}{\bibfnamefont{H.}~\bibnamefont{Xie}},
  \bibinfo{author}{\bibfnamefont{L.}~\bibnamefont{Yu}},
  \bibinfo{author}{\bibfnamefont{X.}~\bibnamefont{Ji}},
  \bibinfo{author}{\bibfnamefont{J.}~\bibnamefont{Li}},
  \bibinfo{author}{\bibfnamefont{Y.}~\bibnamefont{Li}},
  \bibinfo{author}{\bibfnamefont{J.}~\bibnamefont{Dong}},
  \bibinfo{author}{\bibfnamefont{Z.}~\bibnamefont{Shi}}, \bibnamefont{et~al.},
  \bibinfo{journal}{EPL (Europhysics Letters)} \textbf{\bibinfo{volume}{107}},
  \bibinfo{pages}{25001} (\bibinfo{year}{2014}).

\bibitem[{\citenamefont{Shi et~al.}(2019)\citenamefont{Shi, Qiu, Chen, Shi,
  Duan, Xiao, Yang, Zhong, Jiang, Yu et~al.}}]{PShiNF2019}
\bibinfo{author}{\bibfnamefont{P.}~\bibnamefont{Shi}},
  \bibinfo{author}{\bibfnamefont{Z.}~\bibnamefont{Qiu}},
  \bibinfo{author}{\bibfnamefont{W.}~\bibnamefont{Chen}},
  \bibinfo{author}{\bibfnamefont{Z.}~\bibnamefont{Shi}},
  \bibinfo{author}{\bibfnamefont{X.}~\bibnamefont{Duan}},
  \bibinfo{author}{\bibfnamefont{G.}~\bibnamefont{Xiao}},
  \bibinfo{author}{\bibfnamefont{Z.}~\bibnamefont{Yang}},
  \bibinfo{author}{\bibfnamefont{W.}~\bibnamefont{Zhong}},
  \bibinfo{author}{\bibfnamefont{M.}~\bibnamefont{Jiang}},
  \bibinfo{author}{\bibfnamefont{L.}~\bibnamefont{Yu}}, \bibnamefont{et~al.},
  \emph{\bibinfo{title}{Nonlinear coupling induced high frequency axisymmetric
  mode in hl-2a tokamak}},  \bibinfo{journal}{accepted by Nuclear Fusion} (\bibinfo{year}{2019}).

\bibitem[{\citenamefont{Frieman and Chen}(1982)}]{EFriemanPoF1982}
\bibinfo{author}{\bibfnamefont{E.~A.} \bibnamefont{Frieman}} \bibnamefont{and}
  \bibinfo{author}{\bibfnamefont{L.}~\bibnamefont{Chen}},
  \bibinfo{journal}{Physics of Fluids} \textbf{\bibinfo{volume}{25}},
  \bibinfo{pages}{502} (\bibinfo{year}{1982}).

\bibitem[{\citenamefont{Chen et~al.}(2001)\citenamefont{Chen, Lin, White, and
  Zonca}}]{LChenNF2001}
\bibinfo{author}{\bibfnamefont{L.}~\bibnamefont{Chen}},
  \bibinfo{author}{\bibfnamefont{Z.}~\bibnamefont{Lin}},
  \bibinfo{author}{\bibfnamefont{R.~B.} \bibnamefont{White}}, \bibnamefont{and}
  \bibinfo{author}{\bibfnamefont{F.}~\bibnamefont{Zonca}},
  \bibinfo{journal}{Nuclear fusion} \textbf{\bibinfo{volume}{41}},
  \bibinfo{pages}{747} (\bibinfo{year}{2001}).

\bibitem[{\citenamefont{Zonca et~al.}(1996)\citenamefont{Zonca, Chen, and
  Santoro}}]{FZoncaPPCF1996}
\bibinfo{author}{\bibfnamefont{F.}~\bibnamefont{Zonca}},
  \bibinfo{author}{\bibfnamefont{L.}~\bibnamefont{Chen}}, \bibnamefont{and}
  \bibinfo{author}{\bibfnamefont{R.~A.} \bibnamefont{Santoro}},
  \bibinfo{journal}{Plasma physics and controlled fusion}
  \textbf{\bibinfo{volume}{38}}, \bibinfo{pages}{2011} (\bibinfo{year}{1996}).

\bibitem[{\citenamefont{Qiu et~al.}(2013)\citenamefont{Qiu, Chen, and
  Zonca}}]{ZQiuEPL2013}
\bibinfo{author}{\bibfnamefont{Z.}~\bibnamefont{Qiu}},
  \bibinfo{author}{\bibfnamefont{L.}~\bibnamefont{Chen}}, \bibnamefont{and}
  \bibinfo{author}{\bibfnamefont{F.}~\bibnamefont{Zonca}},
  \bibinfo{journal}{Europhysics Letters} \textbf{\bibinfo{volume}{101}},
  \bibinfo{pages}{35001} (\bibinfo{year}{2013}).

\bibitem[{\citenamefont{Smolyakov et~al.}(2008)\citenamefont{Smolyakov, Nguyen,
  and Garbet}}]{ASmolyakovPPCF2008}
\bibinfo{author}{\bibfnamefont{A.}~\bibnamefont{Smolyakov}},
  \bibinfo{author}{\bibfnamefont{C.}~\bibnamefont{Nguyen}}, \bibnamefont{and}
  \bibinfo{author}{\bibfnamefont{X.}~\bibnamefont{Garbet}},
  \bibinfo{journal}{Plasma Physics and Controlled Fusion}
  \textbf{\bibinfo{volume}{50}}, \bibinfo{pages}{115008}
  (\bibinfo{year}{2008}).

\bibitem[{\citenamefont{{Qiu} et~al.}(2014)\citenamefont{{Qiu}, {Chen}, and
  {Zonca}}}]{ZQiuJPSCP2014}
\bibinfo{author}{\bibfnamefont{Z.}~\bibnamefont{{Qiu}}},
  \bibinfo{author}{\bibfnamefont{L.}~\bibnamefont{{Chen}}}, \bibnamefont{and}
  \bibinfo{author}{\bibfnamefont{F.}~\bibnamefont{{Zonca}}},
  \bibinfo{journal}{JPS Conference Proceedings} \textbf{\bibinfo{volume}{1}},
  \bibinfo{eid}{015007} (\bibinfo{year}{2014}).

\end{thebibliography}
\end{document}